\documentclass[%
 reprint,
superscriptaddress,
 amsmath,amssymb,
 aps,
nofootinbib]{revtex4-1}
\usepackage{slashed}
\usepackage{graphicx}% Include figure files
\usepackage{bm}% bold math
\usepackage{amsfonts}
\usepackage{xcolor}

\newcommand{\bwt}{\begin{widetext}}
\newcommand{\ewt}{\end{widetext}}
\newcommand{\newc}{\newcommand}
\newc{\beq}{\begin{equation}}
\newc{\eeq}{\end{equation}}
\newc{\beqa}{\begin{eqnarray}}
\newc{\eeqa}{\end{eqnarray}}
\newc{\nonr}{\nonumber}
\newc{\bi}{\begin{itemize}}
\newc{\ei}{\end{itemize}}
\newc{\ra}{\rightarrow}
\newc{\la}{\leftarrow}
\newc{\lra}{\longrightarrow}
\newc{\lla}{\longleftarrow}
\newc{\Lra}{\Longrightarrow}
\newc{\Lla}{\Longleftarrow}
\newc{\half}{\frac{1}{2}}
\newc{\del}{\delta}
\newc{\Del}{\Delta}
\newc{\eps}{\epsilon}
\newc{\gm}{\gamma}
\newc{\lam}{\lambda}
\newc{\tri}{\triangle}
\newc{\hc}{\dagger}
\newc{\pd}{\partial}
\newc{\wt}{\widetilde}
\newc{\ovl}{\overline}
\newc{\p}{\partial}
\newc{\tchi}{\tilde{\chi}}
\newc{\ds}{\displaystyle}
\newc{\PL}{\hat{L}}
\newc{\PR}{\hat{R}}
\newc{\st}{s_\theta}
\newc{\ct}{c_\theta}

\newc{\clbl}{\color{blue}}
\newc{\clg}{\color{green}}
\newc{\clr}{\color{red}}

\def\nn    {\nonumber}

\begin{document}
\title{Signal for a light singlet scalar at the LHC}% Force line breaks with \\
\author{We-Fu Chang}
\affiliation{%
Department of Physics, National Tsing Hua University, Hsinchu 30013, Taiwan
}%
\affiliation{Theory Group, TRIUMF, Vancouver, B.C. V6T2A3, Canada}

\author{Tanmoy Modak}%
\affiliation{%
Department of Physics, National Taiwan University, Taipei 10617, Taiwan
}%

\author{John N. Ng }
\affiliation{Theory Group, TRIUMF, Vancouver, B.C. V6T2A3, Canada}
\date{\today}

\begin{abstract}

 In the general Higgs portal like models, the extra neutral scalar, $S$, can mix with the Standard Model (SM) Higgs boson, $H$.
We perform an exploratory study focusing on the direct search for such a light singlet $S$ at the Large Hadron Collider (LHC).
After careful study of  the SM background, we find the process $pp\ra t\bar{t}S$ followed by $S\ra b\bar{b}$ can be used to investigate $S$ with mass in the $20<M_S<100$ GeV range, which has not been well explored at the LHC.
The signal significance becomes meaningful with a luminosity around a few $\mbox{ab}^{-1}$.
Also, we study the prospects of finding the light scalar at the future 100 TeV $pp$ collider,
the $Z$ and Higgs factories. With similar luminosity, the current LEP limits on the mixing between $S$ and $H$ can be improved by at least one or two order of magnitudes.

\end{abstract}

\pacs{Valid PACS appear here}% PACS, the Physics and Astronomy
\maketitle

%\tableofcontents
\section{Introduction}
  The discovery of a 125 GeV scalar with properties as expected of the SM
  Higgs boson $H$ gives one last experimental support that it is the correct theory for
  physics at or below the Fermi scale. On the other hand, neutrino oscillations data points to
  massive active neutrinos which will require the model to be amended. Furthermore, the
  SM cannot account for the mounting  evidence for dark matter and neither can it
  explain the matter-antimatter asymmetry of the Universe. Adding SM gauge singlet scalar fields is the
  simplest extension and the most economical and popular scenario to pursue. If one takes only one
  such complex field $S$ and minimally couples it to the SM via interactions with $H$,  the beautiful
  successes of the SM in describing current data are largely undisturbed.

  Theoretically, singlet scalar fields appear in many extensions of the SM.
  They form an essential part of the minimal Majoronic model that accommodates
  both dark radiation\cite{Majoronic DR}and dark matter (DM)
  \cite{Majoronic_DRDM1,Majoronic_DRDM2,Queiroz:2014yna}. The singlet scalar is a good candidate for DM.
  In a different context, singlet scalar extension of the scalar sector helps
  make the electroweak phase transition more first order, which is crucial for a
  successful electroweak baryogenesis\cite{1OPT}. Moreover, the new bosonic degrees of
  freedom contribute to improving the vacuum stability of SM \cite{GLR}\cite{NdP}.

  In this paper we take a simplified model point of view of the singlet Higgs model.
   By that we assume there is only one experimentally accessible real SM singlet scalar which mixes with the SM Higgs.
    It may be part
  of a more complete theory but we only focus on its property as a Higgs portal \cite{PW}. Its only
  interaction with the SM is via couplings to the Higgs field. It may have coupling to other hidden or dark sectors,
  see e.g. \cite{shadow}.
 The production cross section of such light scalar at the collider is entirely determined by its mass and its couplings to the SM fields. Once the scalar is produced, the signal depends on its decay branching ratios. For the collider searches, the relevant parameters are the mass of the scalar, the mixing between the scalar and the SM Higgs, and its decay branching ratios.  Details of how a particular model gives rise to the parameters mentioned above are unimportant for this study.

 To set up our convention, we write the relevant Lagrangian
 adopted from \cite{Majoronic_DRDM1,Majoronic_DRDM2}
 with a SM singlet scalar $S$ as
\beq
\begin{split}
\label{eq:HPL}
{\cal L}\supset &-\mu_H^2(H^\dag H)+  \lambda_H (H^\dag H)^2\\
&  -\mu_S^2 |S|^2 +\lambda_s |S|^4 +\lambda_{SH}(H^\dag H)|S|^2\, .
\end{split}
\eeq
 Spontaneous symmetry breaking (SSB) takes place for both $H$ and $S$. Thus , $\langle S \rangle =v_S/\sqrt{2}$, and
 $\langle H \rangle = v_H/\sqrt{2}$ with $v_H =246$ GeV. The mixing of $H$ and $S$ occurs. Writing
 $ S =(v_S+s_0)/\sqrt{2}$, and $ H =(0, v_H+ h_0)^T/\sqrt{2}$
 the mass square matrix in the basis $\{h_0,s_0\}$ reads
 \beq
 \left(
   \begin{array}{cc}
   2 \lambda_{H} v_H^2 & \lambda_{HS} v_H v_S\\
   \lambda_{HS} v_H v_S &   2 \lambda_{S} v_S^2\\
   \end{array}
 \right)\,.
 \eeq
Denoting the mass eigenstates by
 $H_m,S_m$, we have
\beq
\left(\begin{array}{c}  H_m\\ S_m\end{array} \right)
 = \left( \begin{array}{cc} \ct &-\st \\ \st & \ct
      \end{array} \right)
      \left(\begin{array}{c} h_0\\ s_0  \end{array}  \right)\, .
\eeq
Here we use the shorthand notation $\st,\ct$ for
$\sin\theta, \cos\theta$ and $\theta$ is the mixing angle.

We would like to emphasize again that only $\st$ is relevant for the direct collider searches.
 Our analysis does not depend on the origin of the mixings.
In some other models, the mass square matrix needs not entirely come from SSB.
The mixing satisfies
 \beq
 \tan 2\theta ={ \lambda_{SH} v_H v_S \over \lambda_S v_S^2 -\lambda_H v_H^2 }
 \eeq
 and we set the range of $\theta\in[-\pi/2, \pi/2]$.
 For light singlet, we are interested in the case that $\lambda_S v_S^2 <\lambda_H v_H^2$
and the mass eigenvalues are
 \beqa
 m_{H_m}^2\!=\! \lambda_H\! v_H^2\!+\!\lambda_S\! v_S^2 \!+\!\sqrt{(\lambda_H\! v_H^2\!-\!\lambda_S\! v_S^2)^2\!+\!\lambda_{SH}^2 v_H^2v_S^2}\,,\nonr\\
  m_{S_m}^2\!=\! \lambda_H\! v_H^2\!+\!\lambda_S\! v_S^2 \!-\!\sqrt{(\lambda_H\! v_H^2\!-\!\lambda_S\! v_S^2)^2\!+\!\lambda_{SH}^2 v_H^2v_S^2}\,.
 \eeqa
The convention is set so that  when  $\lambda_{SH}=0$, i.e. no mixing between $H$ and $S$, the above go back to $ m_{H_m}^2= 2\lambda_H v_H^2$ and $ m_{S_m}^2= 2\lambda_S v_S^2$.
$H_m$ is  identified as the 125 GeV SM-like Higgs  boson and $S_m$ is the new neutral scalar boson.
$M_H$ is the mass of $H_m$ which is 125 GeV whereas the mass $M_S$ of $S_m$ is unknown. For
notational simplicity we shall drop the subscript $m$ for the mass eigenstates.

The main goal
of our work is to give a careful study of the constraint on $M_S$ and $\st$ from LHC and future colliders.
Before we embark on this, it is well known  that if the Higgs boson mixes with another scalar, independent of the
mass of the unknown particle, the strength of the Higgs boson to other SM particles will be
multiplied by $\ct$ at the amplitude level. This applies to $S$ with $\ct\ra \st$. The characteristic feature
$S$ shares with the SM Higgs is that their couplings to the massive SM fields are proportional to the masses of the SM particles.
Hence, $t,b,W,Z$ will be good choices for signal detection.
This further implies that all Higgs boson signals currently measured at the LHC will have reduced
SM-like strength. The SM signal strength is parameterized  by $\mu$, and it is unity for the SM.
The latest LHC-1 bound is $\mu=1.09\pm 0.11$ \cite{LHC1_mu} which amounts to $\sin^2\theta<0.13$ at 2$\sigma$\footnote{ At $2\sigma$ C.L., $ 0.87< \mu=\ct^2 <1.31$ implies that $ \st^2<0.13$.  }.
At LHC14, $\sin^2\theta$ can be bounded to be less than $2\times 10^{-2}$ from signal strength\cite{Dawson:2013bba}.
On the other hand, the discovery of $S$ depends crucially on $M_S$.
The obvious mass effect is kinematical, i.e. the heavier $S$ is the lower the production rate at a given energy.
The second effect is in the signals for $S$ detection.  To give an example, for $M_S =400 $ GeV
the dominant decay mode will be  $S\ra WW,ZZ,t\bar{t}$
 if the coupling to the hidden sector is not too large.
In contrast, for a 40 GeV $S$, $\bar{b} b$ will be its dominant decay. Since the final states
signals are drastically different one needs to enhance the signal by optimizing $W,Z$ or $t$ quarks detection for
the heavy $S$ case. For the 40 GeV $S$ a more efficient $b$ quark tagging will help greatly. Another issue
is the SM background suppression. For the example given above it is clear that heavy and light
 scalars will require very different strategies for suppressing the backgrounds.

At the LHC, the dominant production mechanism for the SM Higgs
and the scalar ($S$) that mixed with it is via gluon fusion ( GF).  The subdominant production
processes are the vector boson fusion (VBF), the vector boson associated productions,
 $pp\ra  W^\pm H/ W^\pm S,~ZH/ZS$ (denoted as  $WH~(WS)$ and $ZH~(ZS)$ correspondingly), and
the top-quark pair associated production $pp\ra t\bar{t}H/t\bar{t}S$ (denoted as $t\bar tH/t\bar{t}S$).
The production cross sections of VBF,$WH/WS$, $ZH/ZS$, and $t\bar tS$ are around $10^{-1}-10^{-2}$ times that of the GF.
For the SM Higgs, or the scalar $S$ if it is lighter than 160 GeV, $b\bar{b}$ is the dominant decay modes. However,  the overwhelming
QCD background, mainly from $gg\ra b\bar{b}$, $gg\ra t\bar t$ and single top production ($t\bar{b}$), make the discovery challenging in this
decay mode. Furthermore when the scalar $S$ is heavier than 100 GeV, we have the luxury of using the rather rare but relatively
clean decay modes such as  $S\ra 2\gamma$ or $S\ra ZZ^{(*)}\ra \ell\bar{ \ell}\ell\bar{ \ell}$ for detection.
If the scalar $S$ is heavier than 160 GeV, the decays $WW^{(*)},~ZZ^{(*)}$  provide additional handles for probing.
The limit on the mixing between $S$ and $H$ is a by-product of direct searching for the SM Higgs at the LHC using the aforementioned clean signals \cite{Robens:2015gla,Dupuis:2016fda}.
 The current limit is rather weak, $\st^2<0.25$ for $80<M_S<600$ GeV except that $\st^2<0.16$ when $M_S$ is in the window of $100<M_S<150$ GeV.

However, when $M_S=40$ GeV, the 2-photon decay branching ratio drops to $\sim {\cal O}(10^{-4})$ and the branching ratio
for $S\ra 4 \ell$ drops to $\sim {\cal O}(10^{-7})$. The $b\bar{b}$ decay is the only visible handle left to probe a light $S$ at the LHC.
 Still, the QCD backgrounds are prohibitive for the GF and VBF processes
to be utilized. Moreover, subdominant processes such as $ZS$, $WS$, $t\bar b S$ are not useful for discovery due to
huge $t \bar t$, single top, $W^{\pm}+$ and $Z+$ heavy flavor jets backgrounds.
In order to discover such a scalar $S$ at the LHC, we turn our attention to subdominant production process
$t\bar{t}S$, followed by $S\to b \bar b$ resonance with at least one top decaying leptonically.  The high $b$-jet multiplicity i.e.$ > $ 3 $b$-jets , suppresses
the dominant $t\bar t$ and single top backgrounds to a sufficiently low level, making $t\bar{t}S$ the most promising channel for the discovery of a low mass $S$ at the LHC.
Given the smallness of the $t\bar{t}S$ production cross section, a high luminosity is required. A ballpark estimation is helpful here.
To reach the sensitivity of  $\st^2\lesssim {\cal O}(0.01)$ for the light S,
the required luminosity is around $100\times(\sigma_{GF}/\sigma_{t\bar{t}S})\times(Br(S\ra 2\gamma)/Br(S\ra b\bar{b}))\times 10~\mbox{fb}^{-1}\sim 10^{3}$ fb$^{-1}$.
Such a luminosity is feasible at the high luminosity run of the LHC.

In this paper we focus on the range $20~\text{GeV} < M_S < 100~\text{GeV}$. The lower limit is given by
the energy resolution of LHC.  Furthermore, due to the proximity of many  bottomonium resonances in this energy range, our background analysis will not be applicable and the task of identifying a scalar
resonance in the $b\bar{b}$ mode is nearly impossible. The upper limit is set by our choice of using multiple $b$-quarks as our signal for $S$ detection.
For $M_S>100$GeV, the current LHC Higgs direct search strategies are more powerful than the $t\bar{t}S$ signal.
As the luminosity increased to ${\cal O}(10^3)$ fb$^{-1}$, the current limit on $\st^2$ from direct search is expected to get an order of magnitude improvement as well.
Moreover, searches in this low mass range at the LHC has not been covered by the existing studies on the general Higgs portal models\cite{JWells,Robens:2015gla,
Dupuis:2016fda,Plehn:2009rk,Bock:2010nz,Englert:2011aa,Englert:2011yb,Martin-Lozano:2015dja,Falkowski:2015iwa,MW,Ria:2017qxd}.

We will study the SM background in detail using the current knowledge of $b$-jets
identification as documented by both ATLAS and CMS collaborations. Currently, we have  very limited knowledge
on $S$ in this mass range.  The only experimental information we have is the direct search limit from
LEPII\cite{LEP}. Hence, it is worthwhile to do a careful analysis to see how high luminosity (HL) LHC can improve on this.
We also extrapolate to a future 100 TeV hadron collider. As a comparison, we also study how a Higgs and Z
factory options can shed light on this issue by focusing on a few very clean reactions.

We note that for $M_S< 10$ GeV stringent limits on $\sin^2\theta$ are given by rare $B$ and $K$ mesons
decays \cite{rareBK}. If $M_S<2$GeV, $\sin^2\theta<10^{-6}$ from $B^+\ra K^+ +\mbox{nothing}$.
When $M_S<0.36$ GeV, $\sin^2\theta<10^{-8}$ from $K^+\ra \pi^+ +\mbox{nothing}$. Thus, our study
fills the gap between this and the heavy Higgs searches at the LHC.

The paper is organized as follows. In Sec.II we summarize some relevant details of the simplified
singlet scalar model. Sec. III gives the phenomenology of the model. Details of the LHC study are given.
The chosen signals at the Z and Higgs factory options at the future $e^+ e^-$ colliders are calculated.
The constraints from muon $g-2$ and $B_S\ra \mu\bar{\mu}$ are discussed as well.
A summary and conclusions are given in Sec. IV.

\section{Model}
The Lagrangian for the simplified scalar singlet field is given in Eq.(\ref{eq:HPL}).
The vacuum positivity requires that $\lambda_H,\lambda_S>0$ and $ \lambda_{SH}>-2\sqrt{\lambda_H \lambda_{S}}\,.$
After SSB, one has
\beqa
\mu_H^2 = \lambda_H v_H^2 +\frac{\lambda_{SH}}{2}v_S^2\,,\nonr\\
\mu_S^2 = \lambda_S v_S^2 +\frac{\lambda_{SH}}{2}v_H^2\,.
\eeqa

In terms of $M_H(=125$ GeV), $M_S$, $v_S$, and mixing, the scalar quartic couplings can be expressed as
\beqa
\lambda_H = \frac{1}{2v_H^2}(\ct^2 M_H^2 +\st^2 M_S^2)\,,\nonr\\
\lambda_S = \frac{1}{2v_S^2}(\st^2 M_H^2 +\ct^2 M_S^2)\,,\nonr\\
\lambda_{SH} = -\frac{\st \ct}{v_S v_H}( M_H^2 - M_S^2)\,.
\eeqa

Among the mass eigenstates, the triple scalar coupling vertices  are
\beqa
HHH &:& -3i M_H^2 \left( \frac{\ct^3}{v_H} -\frac{\st^3}{v_S }\right)\,,\nonr\\
HHS &:& -i \st \ct \left( \frac{\ct}{v_H} +\frac{\st}{v_S }\right)(2M_H^2 +M_S^2)\,,\nonr\\
SSH &:& -i \st \ct \left( \frac{\st}{v_H} -\frac{\ct}{v_S }\right)(2M_S^2 +M_H^2)\,,\nonr\\
SSS &:& -3i M_S^2 \left( \frac{\st^3}{v_H} +\frac{\ct^3}{v_S }\right)\,.
\eeqa

In general ( as long as $\st/v_H \neq \ct/v_S$ ), the $H\ra 2S$ decay mode opens up when $M_S<M_H/2$.
Some weak constraints on $\st$ can be derived from the SM Higgs total decay width, see for example\cite{Robens:2015gla,Dupuis:2016fda}.
However, at the LHC, we do not consider the  $pp\ra H\ra 2S$ signal
due to the smallness of the production cross section and the colossal SM background.

Since the quartic scalar couplings  are irrelevant for our study, and they can be easily read from the Lagrangian,
we do not spell them out here.
The modifications to all the other SM Higgs-like couplings are straightforward: each $H/S$ will contribute one extra
power of $\ct/\st$ suppression factor. Note that the mixing is only between the two scalar parts, the Feynman rules for the Goldstone bosons remain unchanged.

When $M_S>2 m_b$, the dominating visible decay is $S\ra b\bar{b}$. The partial width is given by
\beq
\Gamma_{S\ra b\bar{b}} = \frac{\st^2 N_c M_S }{8\pi}\left(\frac{m_b}{v_H}\right)^2 \left(1-4\frac{m_b^2}{M_S^2}\right)^{3/2}\,.
\eeq
Taking into account the LEP limit on $\st$\cite{LEP}, the partial width is small, $\lesssim 10^{-4}\mbox{GeV}$ for $M_S<80$ GeV, and makes $S$ a very narrow resonance for discovery. There can be additional decays into invisible modes if S has a coupling to the
dark sector. This will be model dependent.

%%%%%%%%%%%%%%%%%%%%%%%%%%%%%%%%%%%%%%%%
\section{Phenomenology}
%%%%%%%%%%%%%%%%%%%%%%%%%%%%5
\subsection{ LHC signal}
%%%%%%%%%%%%%%%%%%%%%%%%%%%%%
In  this subsection we investigate the discovery potential of light $S$ at the LHC via $pp$ collision.
The primary production mechanism of $S$ at LHC is via $pp\to t \bar{t}S$, followed by
$S \to b \bar{b}$ decay and leptonic decays of at least one of the top quarks; constituting
four $b$-tagged jets, at least one charged lepton plus missing ($E_T^{\mbox{miss}}$) signature.
The final state topology is similar to SM $t\bar t H$ (with $H\to b\bar b$) production~\cite{ATLAS:2016awy},
which we closely follow in our analysis. The signal region will have subdominant contributions from single top productions such as
$pp\to t S \bar{b}/j$ (conjugate processes implied).
We also include several other processes namely:  $p p \to \bar{t} W^+ S$, $p p \to W^+ S H$, $p p \to W^+ S b \bar{b}$ ,
$p p \to W^+ Z S$ and $p p \to Z Z S$ etc., which are possible within the model and provide small contributions to the signal region.
There exists several sources SM backgrounds, such as $t\bar{t}$+heavy flavor jets ($t\bar{t}$+h.f. jets),
$t\bar{t}$+light flavor jets ($t\bar{t}$+l.f. jets)with subdominant contributions from $t$- and $s$-channel single top
production with contributions from $t\bar{t}Z$, $t\bar{t}H$, $tWH$ and other processes such as $Z/\gamma^*+4b$-jets.
We have not included non-prompt and fake backgrounds in our analysis. These contributions are not
properly modeled in Monte Carlo event generators and one needs data to estimate such contributions.

The signal and background samples are generated at leading order (LO) in Monte Carlo event
generator MadGraph5\_aMC@NLO~\cite{Alwall:2014hca} adopting NN23LO1 PDF set~\cite{Ball:2013hta}, interfaced with
Pythia~6.4~\cite{Sjostrand:2006za} for showering, hadronization and underlying events.
The Matrix element (ME) of the signal and background processes are generated up to one additional jet in
the final state except for $t\bar{t}$+heavy flavor jets. The ME for the latter is generated up to two additional jets in the final state.
We follow  MLM~~\cite{Alwall:2007fs} matching prescription for the ME and parton shower (PS) merging. Due to computational limitations we
restrict the number of additional jets in ME to the above mentioned numbers for the respective signal and background processes.
Events are finally fed into Delphes~3.3.3~\cite{deFavereau:2013fsa} for fast detector simulation (ATLAS based).
The rejection factor for light jets and charm jets are assumed to be 1/137 and 1/5 respectively~\cite{ATLAS:2014ffa}.
The jets are reconstructed using anti-$k_t$ jet algorithm with radius parameter $R=0.5$.
The effective model is implemented using FeynRules~2.0~\cite{Alloul:2013bka}.

\textbf{Selection cuts}: Events are selected such that it should contain at least
four jets out of which at least four are $b$-tagged and at least one charged leptons ($\equiv e, \mu$)
and missing transverse energy $E^{miss}_{T}$ (denoted as $(4j,4b,1\ell)$ process). The $p_T$ of the leading lepton
in an event is required to be $> 25$ GeV, while if it contains a subleading lepton, the
transverse momenta of the subleading lepton is required $> 15$ GeV.
The maximum pseudo-rapidity of the lepton(s) in an event should be $|\eta| < 2.5$.
Minimum $p_T$ of all jets or  b-tagged jets in an event are required to be $>20$ GeV with
pseudo-rapidity $|\eta| < 2.5$. The minimum separation between a jet (or b-jet)
and charged lepton(s) is required to be $\Delta R_{j(b)\ell}> 0.4$,  while the minimum separation
between two charged leptons should be $\Delta R_{\ell\ell} > 0.4$, with missing transverse $E^{miss}_{T} > 35$ GeV.
Events containing hardronic tau lepton with $p_T> 25$ GeV are vetoed.
Each event will have multiple combinations of $m_{bb}$ due to high $b$ jet multiplicity.
The pairs with $m_{bb}$ closest to $M_S$ are assumed to be coming from decay of $S$,
and passed through invariant mass cut $|M_{bb} - M_{S}| < 10~\mbox{GeV}$.

In order to illustrate the discovery potential, we consider six benchmark configurations  in the
minimal Majoronic model for  dark radiation and dark matter\cite{Majoronic_DRDM1,Majoronic_DRDM2}  given as follows:
\begin{align}
&\mbox{Config-1}: v_S=1~\mbox{TeV},~s_\theta= \pm 0.1\,, \nn\\
&\mbox{Config-2}: v_S=1~\mbox{TeV},~s_\theta= \pm 0.3\,, \nn\\
&\mbox{Config-3}: v_S=10~\mbox{TeV},~s_\theta=\pm 0.1\,, \nn\\
&\mbox{Config-4}: v_S=10~\mbox{TeV},~s_\theta=\pm 0.3\,, \nn\\
&\mbox{Config-5}: v_S=100~\mbox{TeV},~s_\theta=\pm 0.1\,, \nn\\
&\mbox{Config-6}: v_S=100~\mbox{TeV},~s_\theta=\pm 0.3\,. \label{eq:conf}
\end{align}

Once $v_S$, $\st$, and $M_S$ are given,  the invisible decay branching ratio and $Br(S\ra b\bar{b})$ are fixed in this particular model.
We find the result is not sensitive to the sign of $\st$, namely, the triple and quartic scalar contributions are negligible.

The signal cross sections for different configurations after selection cuts are plotted in Fig.~\ref{cross},
while the background cross sections are given in the appendix for $\sqrt{s}= 13~\mbox{and}~100$ TeV
respectively.
To estimate signal significance, we use the following formula~\cite{Cowan:2010js}:
\begin{align}
\mathcal{Z} = \sqrt{2 \left[ (S+B)\ln\left( 1+\frac{S}{B}\right)-S \right]}\,, \label{eq:signif}
\end{align}
where $S$ and $B$ are the number of signal and background events respectively.
We require $\mathcal{Z}= 5$ for the 5$\sigma$ discovery and $\mathcal{Z}= 3$ for  $3\sigma$ significance.

 \begin{figure*}[htb!]
 \centering
     \includegraphics[width=0.7\textwidth]{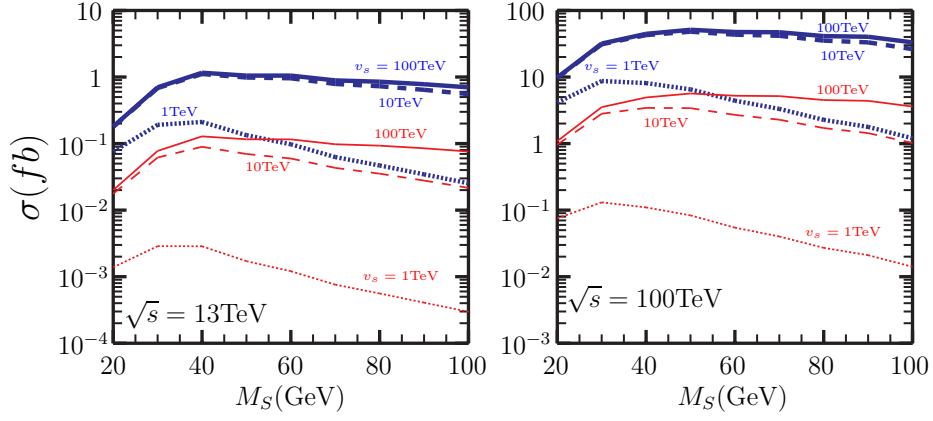}
 \caption{Signal cross sections (fb) for different $M_S$ at 13 (left) and 100 TeV (right) $pp$ collision
 for six different benchmark configurations as given in Eq.\eqref{eq:conf}. The blue/red curves correspond to $|\st|=0.3/0.1$.}
 \label{cross}
 \end{figure*}
 \begin{figure*}[htb!]
 \centering
      \includegraphics[width=0.7\textwidth]{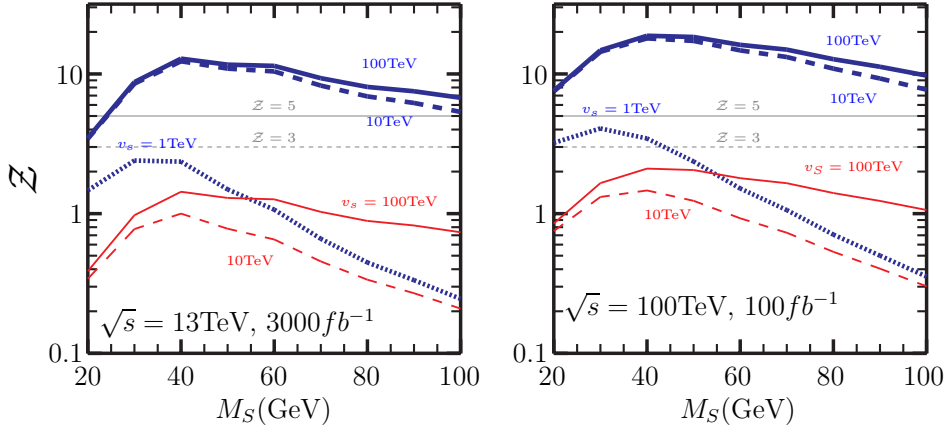}
 \caption{The left panel conforms significance for different benchmark configurations at 13
 TeV LHC with 3000 fb$^{-1}$ integrated luminosity, while the right panel represents the significance
 for  100 TeV $pp$ machine with 100 fb$^{-1}$ integrated luminosity. Color schemes are the same as Fig.~\ref{cross}.}
 \label{signi}
 \end{figure*}

Let us have a closer look at Fig.~\ref{cross} and Fig.~\ref{signi}. In Fig.~\ref{cross} we have plotted the cross section
of the signal for different values of $M_S$ adopting several benchmark scenarios as given in Eq.~\eqref{eq:conf}.
The cross section contours for all the configurations increase initially and reach to the maximum value at $M_S\sim 40$ GeV, finally falling
slowly towards a higher mass. The minimum $p_T$ cuts on the $b$-jets are too strong for $M_S$ below 40 GeV, causing the initial dip
in the cross sections; which finally reach its maximum value $M_S\sim 40$ GeV and then falls again due to the drop in parton luminosity for heavier
$M_S$. The cross section of the signal process $pp \to t\bar t S$ is directly proportional to $\st^2$; hence,
configurations with larger $|\st|$ have larger cross sections resulting in better discovery
probability. However for fixed $s_\theta$ and $M_S$, the cross sections primarily depend on the
$\mathcal{B}r(S\to b \bar b)$. For the configurations with larger $v_S$, the total decay width of $S$ is dominated by $\Gamma_{S\to b \bar b}$
~\footnote{For the $M_S$ value under consideration, the total decay width of
$S$ is nicely approximated as $\Gamma_{tot}\approx\Gamma_{S\to b \bar b} +\Gamma_{S\to c \bar c}+\Gamma_{S\to \omega \bar \omega}$,
 where $\omega$ is the singlet Majoron\cite{Majoronic_DRDM1,Majoronic_DRDM2}. }.
This is evident from the position of the dotted and dashed lines in Fig.~\ref{cross}. Furthermore, the blue solid  and dashed lines
almost coincide due to  $\mathcal{B}r(S\to b \bar b)\approx 1$ for $v_S =10,100$ TeV.
However the red solid and dashed lines are more separated than the blue ones,
primarily due to $\mathcal{B}r(S\to b \bar b)$ being lower for $v_S =10$ TeV than $v_S =100$ TeV.

\begin{figure}
    \centering
    \includegraphics[width=0.45\textwidth]{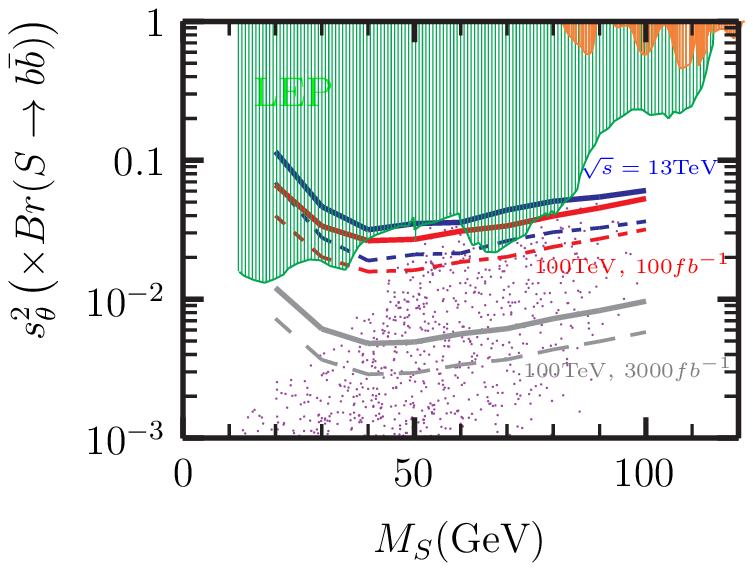}
    \caption{$\sin^2\theta\left(\times Br(S\ra b\bar{b})\right)$ vs $M_S$.
     The light green area corresponds to the LEP exclusion bound\cite{LEP}.
    The orange shade is the exclusion region from the LHC direct search\cite{Robens:2015gla,Dupuis:2016fda} which is independent of $Br(S\ra b\bar{b})$.
     The blue/red curve is for the $pp$ collider with $\sqrt{s}=13/100$ TeV and integrated luminosity $=3000/100~\mbox{fb}^{-1}$, respectively.
       The solid and dashed lines represent $\mathcal{Z}=5,3$ limit respectively.
     The gray curves are the evidence and discovery limits at the $100$ TeV $pp$ collider with $3000~\mbox{fb}^{-1}$ integrated luminosity.
       The scattered points are the realistic configurations found in \cite{Majoronic_DRDM2}.}
    \label{fig:sum_PP_fig}
\end{figure}

We restrict ourselves in the mass range $M_S=[20,100]$ GeV. For $M_S < 20$ GeV the controlling of the background would be much
more difficult due to the presence of bottomonium resonances, while $M_S > 100$ GeV  the dominant signal process suffers interference with the
SM $t\bar t H$ process. The strong $|M_{bb} - M_{S}| < 10$ GeV cut is primarily used to minimize the interference with the SM
$t \bar t H$ process near $m_S \sim 100$ GeV. For simplicity we use the same invariant mass cut even for the lower $M_S$ region.
Note that we have not included QCD corrections to signal and SM background processes, which may induce some uncertainties in our
results. Since our study is exploratory it is premature to include
these higher order effects.
Together with all the current limits, our findings are summarized in Fig.\ref{fig:sum_PP_fig}.
It is encouraging to see that despite the large SM background, HL-LHC13 can improve upon
LEPII results on $\st^2$.

 We naively extrapolate our study to the $100$ TeV $pp$-collider.
The cross sections for both the signal and the background are enhanced due to the rise of gluonic parton distribution at small $x$.
We find the signal significance in this case with $100~\mbox{fb}^{-1}$ integrated luminosity is slightly better than that at the HL-LHC, which may alleviate due pile-up effects. Such effects are included in a realistic experimental analysis, which we do not incorporate
in our exploratory study.

\subsection{ Signal at the Z factory}

In the planned Z-factory option of FCC-ee , $\sim{\cal O}(10^{12})$  Z bosons are expected each year\cite{TLEP}.
This opens up a new avenue to directly probe the light scalar with the rare Z decays.
The signal will be $e^+e^-\ra Z \ra S f\bar{f}$, and $S$ subsequently decays into $b\bar{b}$ pair.
A useful kinetic variable $y_b\equiv m^2_{bb}/M_Z^2$  is defined where $m_{bb}$ is the invariant mass of the $b\bar{b}$ pair.
The on-shell light scalar  gives a very narrow resonance peak in $y_b$ at around $y_b=(M_S/M_Z)^2$ and stands out from the continuous SM background.
The signal branching ratio is given by\cite{H_hunter}
\beq
Br(Z\ra S f\bar{f})= \st^2 \times F(M_S/M_Z) \times Br(Z\ra f\bar{f})\,,
\eeq
where
\beqa
F(r)=\frac{G_F M_W^2 }{24\sqrt{2} \pi^2 c^2_W}
\left[ {3 r (r^4-8r^2+20)\over \sqrt{4-r^2}}\cos^{-1}\left( \frac{r}{2}(3-r^2)\right)\right.\nonr\\
\left.-3(r^4-6r^2+4)\ln r-\frac{1}{2}(1-r^2)(2r^4-13r^2+47) \right]\,.\nonr
\eeqa
Here $c_W(s_W)$ is the shorthand for the weak mixing $\cos\theta_W(\sin\theta_W)$.
One of the signals we are interested in is $Z\ra \nu\bar{\nu}S;S\ra b\bar{b}$,  and the SM background is
the 4-body decay $Z\ra b\bar{b} \nu\bar{\nu}$.
To quantify the discovery potential at the future Z-factory,
we calculate the significant $\mathcal{Z}$.
The signal will be
\beq
S= \st^2 \times F(r_S)\times Br(Z\ra \nu\bar{\nu})\times Br(S\ra b\bar{b})\times N_Z\,,
\eeq
 where $r_S=(M_S/M_Z)$,  $N_Z$ is the fiducial number of Z-bosons, and we sum over all neutrino species.
 We use CalcHep\cite{calcHEP} to calculate  numerically the SM differential decay width, $d\Gamma^{SM}(Z\ra b\bar{b}\nu\bar{\nu})/d y_b$
 ( see Fig.10 in \cite{Majoronic_DRDM2}).
Then the signal background is obtained by integration over the continuous distribution in the vicinity of $y_b=r_S^2$
\beq
\mbox{B}= N_Z\times \int^{r_S^2+\delta y_b}_{r_S^2-\delta y_b} dy_b { d\Gamma^{SM}(Z\ra b\bar{b}\nu\bar{\nu}) \over d y_b}\,.
\eeq
 Assuming the invariant mass resolution at the FCC-ee is 1GeV, then $\delta y_b$ is roughly $ 2 \frac{M_S \times 1 \mbox{GeV}}{M_Z^2} \sim \frac{r_S}{45}$.
Taking $N_Z=10^{12}$ and $Br(Z\ra \nu\bar{\nu}) =0.2000(6)$\cite{PDG2016}, the curves of $\mathcal{Z}=3, 5$ are displayed in Fig.\ref{fig:sum_EE_fig}.
 Not surprisingly, at a Z-factory one could probe the new light scalar in the mass range from 20 to 80 GeV with
 mixing orders of magnitude smaller than  the current LEP bound and far better than a high energy
 pp collider.

\subsection{ Signal at the Higgs factory}
\begin{figure}
    \centering
    \includegraphics[width=0.4\textwidth]{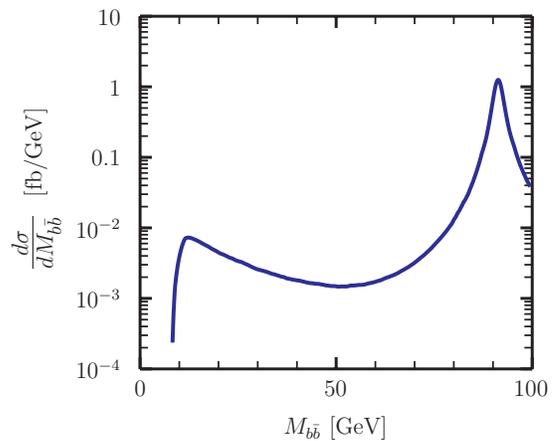}%\put(230,-15){$M_S(GeV)$}
    \caption{The SM background for $e^+e^-\ra Z S$ at the Higgs factory with $\sqrt{s}=240$GeV. See the text for details. }
    \label{fig:SMBG_HF}
\end{figure}

The plan for Higgs factory is an $e^+ e^-$ collider operating at $\sqrt{s}=240-250$ GeV and luminosity of a few $\mbox{ab}^{-1}$.
For definiteness we take $\sqrt{s}=240$ GeV and,  to take into account the
detection efficiency,  use an integrated fiducial luminosity $\mathfrak{L}=1 \, \mbox{ab}^{-1}$ as the benchmark.
The tree-level scattering cross section is
\beq
\sigma(e^+e^-\ra Z S) ={\st^2 \pi \alpha^2 \over 3 s_W^4 c_W^4 }\left(g_{eL}^2+g_{eR}^2\right)
 { p_{cm}(p_{cm}^2 +3 M_Z^2)\over \sqrt{s}(s-M_Z^2)^2}
\eeq
where $g_{eL}=s_W^2-\frac12$, $g_{eR}=s_W^2$,
and $p_{cm}=\sqrt{(s+M_S^2-M_Z^2)^2/(4s)-M_S^2}$ is the CM momentum of the $S$.
To be more specific, we consider the final states $Z\ra \mu\bar{\mu}$ and $S\ra b\bar{b}$ as an example.
Then the number of signal events will be:
\beq
S=  \mathfrak{L}\times \sigma(e^+e^-\ra Z S)\times Br(S\ra b\bar{b})\times Br(Z\ra \mu\bar{\mu})
\eeq
and $ Br(Z\ra \mu\bar{\mu})=3.366(7)\%$\cite{PDG2016}.
The SM background is dominated by the t-channel $e^+e^-\ra Z+Z$ diagram and the complete tree-level contribution is
evaluated numerically by CalcHEP. We applied two cuts: (1)
$|\,M_{\mu\bar{\mu}}-M_Z\,|<1$ GeV,  and (2)$|\,E_{\mu\bar{\mu}}-\sqrt{M_Z^2+p_{cm}^2}\,|<1$ GeV to suppress the SM background.
See Fig.\ref{fig:SMBG_HF} for the result. One can clearly see the huge
peak around $M_Z$ due to the $Z$ resonance, and the photon contribution accounts
for the rise at the low energy end.

\begin{figure}
    \centering
    \includegraphics[width=0.45\textwidth]{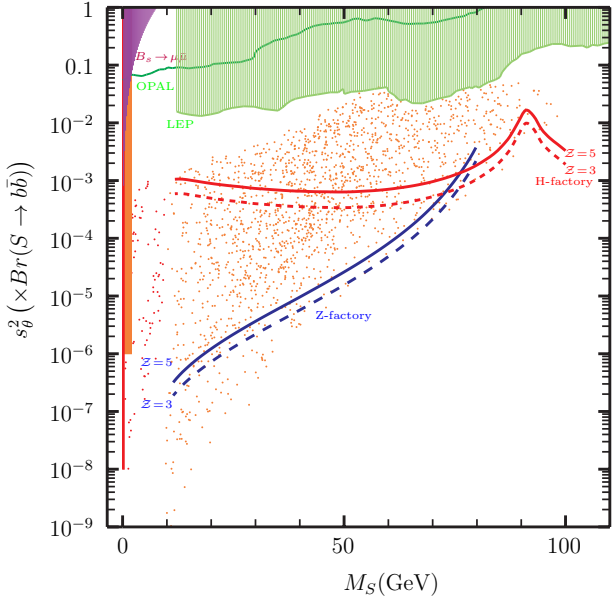}%\put(230,-15){$M_S(GeV)$}
    \caption{$\sin^2\theta\left(\times Br(S\ra b\bar{b})\right)$ vs $M_S$.
     The left most two are from rare B and K decays. The upper-left region in purple is
    the two sigma exclusion from the $B_s\ra \mu\bar{\mu}$.
    The light green area corresponds to the LEP exclusion bound\cite{LEP}.
    The dark green curve is the decay mode independent bound from OPAL\cite{OPAL}.
    The scattered points are the realistic configurations found in \cite{Majoronic_DRDM2}.
     The blue(red) dashed/ solid curve is for signal significant $\mathcal{Z}=3, 5$ at the Z(Higgs) factory, respectively.
    }
    \label{fig:sum_EE_fig}
\end{figure}

Due to the finite energy resolution, we take $\delta m_{b\bar{b}}=1$ GeV,
the number of SM background events is given by
\beq
B= \mathfrak{L}\times \int^{m_{b\bar{b}} +\delta m}_{m_{b\bar{b}} -\delta m} \frac{\sigma_{SM}(Z\ra \mu\bar{\mu}b\bar{b})}{d m_{b\bar{b}}}\,.
\eeq
The curves for signal significance $\mathcal{Z}=3,5$ are shown in Fig.\ref{fig:sum_EE_fig} along with  other results depicted in Fig.(\ref{fig:sum_PP_fig}).
It can be seen that the sensitivity to $\st^2\cdot Br$ is poorer when $M_S$ is around the Z-pole due
to the large intrinsic SM background. However, the shear statistics make the sensitivity roughly
two orders of magnitude better than that at the LEP.

\subsection{ muon (g-2)}
The current status of the  muon anomalous magnetic moment is $a_\mu^{exp}-a_\mu^{SM}=2.88(80)\times 10^{-9}$\cite{PDG2016}.
Both the modified SM Higgs coupling and the presence of the light scalar contribute to  $\triangle a_\mu$
\beq
\triangle a_\mu = \st^2 {G_F m_\mu^2 \over 4\pi^2\sqrt{2}} \left[G(M_S^2/m_\mu^2)-G(M_H^2/m_\mu^2)\right]
\eeq
at 1-loop level, where
\beq
G(x)=\int^1_0 dy {y^2(2-y) \over y^2+x(1-y)}\,.
\eeq
Note that $G(0)=3/2$; therefore, $\triangle a_\mu \simeq 3.49\times \st^2\times 10^{-9}$ when $M_S\ll m_\mu$.
The modification moves up the SM prediction and alleviates the tension between theory and experiments.
More precisely, to explain the discrepancy at the two sigma lower bound, the mixing has to be
\beq
\st^2={0.5447\over G(M_S^2/m_\mu^2)-9.171\times10^{-6}}\,.
\eeq
The above equation has solution only if $M_S<0.1 $GeV and $\st^2 \gtrsim 0.5$
which has been ruled out by the $K^+\ra \pi^++\mbox{nothing}$.
 Although the light singlet scalar provides a positive contribution to $a_\mu$, it alone cannot accommodate the current discrepancy between theory and experiment.

\subsection{ $B_s \ra \mu \bar{\mu}$}
The presence of the light scalar also contributes to the rare process $B_s \ra \mu \bar{\mu}$.
This transition is governed by the effective Hamiltonian
\beq
{\cal H}_{eff}= -\frac{G_F}{\sqrt{2}}\frac{\alpha}{\pi s_W^2}V_{tb} V^*_{ts}
\left(C_{10}\hat{O}_{10}+C_{S}\hat{O}_{S}+C_{P}\hat{O}_{P}\right)
\eeq
where
\beqa
\hat{O}_{10} =\left(\bar{s}\gamma_\mu \hat{L} b\right)\left(\bar{\mu}\gamma^\mu\gamma_5 \mu\right)\,,\nonr\\
\hat{O}_{S} =\frac{m_\mu m_b}{M_W^2}\left(\bar{s}\hat{R} b\right)\left(\bar{\mu}\mu\right)\,,\nonr\\
\hat{O}_{P} =\frac{m_\mu m_b}{M_W^2}\left(\bar{s}\hat{R} b\right)\left(\bar{\mu}\gamma_5\mu\right)\,,
\eeqa
where $\hat{R} = \frac{1+\gamma_5}{2}$ and $\hat{L} = \frac{1-\gamma_5}{2}$.
Note that the vector part of lepton current in $\hat{O}_{10}$
does not contribute in this process when contracted with the meson momentum.
In SM, the Z-penguin or $\hat{O}_{10}$ dominates the decay.
The complete 1-loop expressions of the Wilson coefficients,
$C_S$ and $C_P$, can be found in \cite{CS_CP}, and we do not repeat that here.
The key is that the SM  Higgs gives a negative contribution $- 3 m_t /8M_H$ to $C_S^{SM}(\simeq -0.939)$.
The branching ratio is given by
\beqa
Br(B_s \ra \mu \bar{\mu})= { \tau_{B_s} G_F^4 M_W^4\over 8\pi^5} |V_{tb} V^*_{ts}|^2f^2_{B_s}M_{B_S}m_\mu^2\nonr\\
\times
\sqrt{1-4 m_\mu^2/M_{B_s}^2} (|\mathbf{P}|^2+|\mathbf{S}|^2)
\eeqa
where
\beqa
\mathbf{P}=C_{10}+ \frac{M^2_{B_s}}{2M_W^2}{m_b\over m_b+m_s} C_P\,,\nonr\\
\mathbf{S}=\sqrt{1-4 m_\mu^2/M_{B_s}^2}\frac{M^2_{B_s}}{2M_W^2}{m_b\over m_b+m_s} C_S\,.
\eeqa
The reduction of the SM Higgs coupling and the light scalar modify the
scalar penguin contribution to this decay, and
\beq
C_S^{SM} \Rightarrow C_S^{SM} -\frac{3}{8}\st^2\left(\frac{m_t^2}{M_S^2}-\frac{m_t^2}{M_H^2} \right)\,.
\eeq
When $M_S< M_H$, $|C_S|>|C_S^{SM}|$.
On the other hand, $C_{10}$ and $C_P$ remain the same.
Therefore, the $B_s\ra \mu\bar{\mu}$ branching ratio is larger than
the SM one in the general Higgs portal model with a light singlet scalar.
What experimentally measured is the time averaged decay branching ration
$\overline{Br}(B_s \ra \mu \bar{\mu})$ due to the sizable $B_s-\overline{B_s}$ oscillation\cite{R_B_time_ave}.
The current experimental result agrees with the SM prediction\cite{R_B_s_2017}
\beq
\overline{R}= \frac{\overline{Br}(B_s \ra \mu \bar{\mu})_{exp}}{\overline{Br}(B_s \ra \mu \bar{\mu})_{SM}}=0.84\pm0.16\,,
\eeq
where the results measured at LHCb\cite{LHCb_B_s2017} and CMS\cite{CMS_B_s} and the SM prediction from\cite{B_s_SM} are used.

Numerically, at two sigma ($\overline{R}<1.16$), we find
\beq
\st^2 <{496.06 \over m_t^2/M_S^2-1.91}
\eeq
for $M_S<7.76$GeV. That the mixing can be constrained is understandable since
the negative scalar penguin contribution becomes important and increases the branching ratio when the
second scalar is light.
 However, the constraint on $\st^2$ from $B_s\ra \mu\bar{\mu}$ for $M_S<7.76$GeV
cannot compete with the limit given by the decay-mode independent searches at LEP\cite{OPAL}.

\section{Summary}

We have analyzed the mixing of a singlet scalar with the SM Higgs boson in the challenging mass range
of 10 to 100 GeV. We found that using the signal of $ 4 b\text{-jets}+ \geq 1\ell+ \text{MET}$ one can
successfully probe this mass range with limits on the mixing angle $\sim$ an order better than the
current LEPII results. We have used very conservative cuts and search criteria established by ATLAS and
CMS collaborations. We did not use more advanced techniques such as boosted t-quarks nor neural
networks which will improve  our exploratory study. The significance of our results only applies
to HL-LHC due to the smallness of the cross section for $pp\ra t\bar{t}S$. An order of magnitude
improvement over the LEPII results is achievable at LHC13 (see Fig.\ref{fig:sum_PP_fig})
for $80~\text{GeV}< M_S< 110~\text{GeV}$.
This analysis is then naively extended to a future 100 TeV pp collider.
Our calculation shows that not much is gained if the luminosity is relatively low.
In contrast, a high luminosity 100 TeV collider can improve on
the LEPII limit by up to two orders of magnitude in a mixing angle scan. We realize that such a futuristic
machine will have different systematics and experimental challenges and theoretical uncertainties
that make such an extrapolation speculative at best. Nonetheless, we find the result intriguing.

The discovery potential of the 125 GeV Higgs in the $t\bar t H$ production with $H \to b \bar b$ decay has been
extensively discussed in Refs.~\cite{Plehn:2009rk,Buckley:2015vsa,Li:2015kaa,Moretti:2015vaa}.
With the collection of 3 ab$^{-1}$ data, the $t\bar t H$ signal strength can be probed up to $\sim20\%$
level of the SM prediction~\cite{Moretti:2015vaa}. If measured, a lower
deviation of the $t\bar t H$ signal strength from the SM prediction can be attributed to the mixing
between the singlet scalar $S$ and 125GeV $H$. However, the mass of $S$ can not be ascertain from this alone. Thus, such a deviation
will act as an indirect probe for the singlet $S$ in our model. Clearly, a direct probe via $t\bar t S$, $S\to b \bar b$ would be essential.
We also remark that a search for light CP-even resonance in association with $t\bar t$ production,
based on simplified model has been discussed in \cite{Dolan:2016qvg}.

The signal for light singlet scalar discovery we studied hinges on a sizable $b\bar{b}$ decay.
For models which have large portal Higgs to dark sector coupling (HD), our results can be scaled
by scaling the $b\bar{b}$ branching ratio. If the HD is large, then an efficient  missing energy
signal will have to be used. However, we have not found a signal in which $S\ra \text{invisible}$ can be
detected above the SM background at the LHC for this range of $M_S$.

As a comparison we also studied how the Z-factory and Higgs factory options for a future $e^+ e^-$ collider
can be used to study the light scalars. We found that the $Z\ra S\nu\bar{\nu};S \ra b\bar{b}$ is the most promising signal to
search for. If not found, the limit set on the mixing angle will be 3-4 orders of magnitude better
than LEPII.
We also note that due to the stringent bound from $K\ra\pi \nu\nu$, the  muon g-2 discrepancy cannot be explained alone by the presence of very light scalar.
 Moreover, the recent $B_s\ra\mu\bar{\mu}$ data does not provide a better constraint on $\st^2$  for $2<M_S<7.7$ GeV than that from the branching ratio independent light scalar search at LEP~\cite{OPAL}.

\begin{acknowledgments}
 WFC is supported by the Taiwan Minister of Science and Technology under
Grant No. 106-2112-M-007-009-MY3  and No.105-2112-M-007-029.
TM is supported by Ministry of Science and Technology of
R.O.C. under the Grant No. MOST-106-2811-M-002-187. TM thanks Stathes Paganis
for discussions.
\end{acknowledgments}

\section{appendix }
The cross sections of the different background processes of the $(4j,4b,1\ell)$ process for different $M_S$,
with $\sqrt{s}= 13$ and 100 TeV $pp$ collision are
presented in Table~\ref{bkg2_13} and Table~\ref{bkg2_100} respectively.
\begin{table}[htbp!]
\centering
\begin{tabular}{|c|c|c|c|c|c|c|c|c|c|c|c|c|c|}
\hline
&&&&&&&&\\
 $M_S$         & $t\bar{t}$+h.f. & $t\bar{t}$+l.f.  & Single   & $t\bar{t}Z$ & $t\bar{t}H$ & $tWH$      & Oth-  & Total  \\
 (GeV)       &   jets            &  jets            &  top       &              &             &          & ers        & BG      \\
\hline
\hline

\hline
  20                &  7.75               &  0.116            &0.005          & 0.009      &  0.035     & 0.003   & 0.005   & 7.923   \\
% &&&&&&&&\\
  30               & 18.41               & 0.321             &  0.013       & 0.03        &0.148        &0.009    & 0.032   & 18.936   \\
% &&&&&&&&\\

  40                & 23.26               &  0.486            &  0.018        & 0.074      & 0.279    &0.013    & 0.042   & 24.172   \\
% &&&&&&&&\\
  50               &  22.29              & 0.605             &  0.02        & 0.126      & 0.418      & 0.019    & 0.045    & 23.523   \\
% &&&&&&&&\\
  60                &  23.26              & 0.706             &   0.023       & 0.200      & 0.501    & 0.022    &  0.059  &  24.771  \\
% &&&&&&&&\\
  70               &  25.20              &  0.839            &   0.024       & 0.266      & 0.590     & 0.024    & 0.076   &  27.019  \\
% &&&&&&&&\\
  80                & 31.01              &  0.911            &  0.026        & 0.316      & 0.665     & 0.029    & 0.081   &  33.038  \\
% &&&&&&&&\
  90                 &  30.04             &  0.912           & 0.026          & 0.326      & 0.711    &  0.032   & 0.085   & 32.132   \\
% &&&&&&&&\\
  100              & 30.04             &  0.922           & 0.025         & 0.292       &0.787        & 0.036    & 0.081   & 32.183   \\
% &&&&&&&&\\

\hline
\end{tabular}
\caption{The cross section in fb of different background processes for after
applying selection cuts. The cross section is estimated at $\sqrt{s}= 13$ TeV LHC.}
\label{bkg2_13}
\end{table}
\begin{table}[htbp!]
\centering
\begin{tabular}{|c|c|c|c|c|c|c|c|c|c|c|c|c|c|}
\hline
&&&&&&&&\\
$M_S$         & $t\bar{t}$+h.f. & $t\bar{t}$+l.f.  & Single   & $t\bar{t}Z$ & $t\bar{t}H$ & $tWH$         & Oth-  & Total  \\
 (GeV)       &   jets            &  jets            &  top       &              &             &            & ers        & BG      \\
\hline
\hline

\hline
  20         &  156.02              & 4.90           & 0.10              & 0.48       & 2.20       &  0.17     & 0.1   & 163.97     \\
% &&&&&&&&\\
  30         &  429.04              & 13.47          & 0.26              & 1.57       & 8.36       &  0.50      & 0.69  & 453.89     \\
% &&&&&&&&\\
  40         &  507.05              & 20.36          & 0.38              & 3.98       & 15.33      &  0.79      & 1.29  & 549.18       \\
% &&&&&&&&\\
  50         &  702.07              & 24.96          & 0.43              & 7.27       & 21.47      &  1.05      & 1.39  & 758.64     \\
% &&&&&&&&\\
  60         &  780.08              & 28.89          &  0.46             & 9.28       & 26.72      &  1.31      & 1.29  & 848.03    \\
% &&&&&&&&\\
  70         &  897.08              & 33.83          &  0.49             & 13.51      & 30.45      &  1.52      & 1.57  & 978.45   \\
% &&&&&&&&\\
  80         &  936.09              & 36.29          &  0.51             & 17.97      & 35.15      &  1.67      & 2.48  & 1030.16     \\
% &&&&&&&&\\
  90         &  1170.11             & 35.75          &  0.51             & 16.64      & 39.42      &  1.97      & 1.98  & 1266.38      \\
% &&&&&&&&\\
  100        &  1053.10             & 35.18          &  0.51             & 13.75      & 42.95      &  2.23      & 1.78  & 1149.50       \\
% &&&&&&&&\\
\hline
\end{tabular}
\caption{Same as Table~\ref{bkg2_13} but for $\sqrt{s}= 100$ TeV LHC.}
\label{bkg2_100}
\end{table}

\end{document}